# Drone Controller Localization Based on TDoA

Yuhong Wang, Yonghong Zeng, Peng Hui Tan, Sumei Sun and Yugang Ma
Institute for Infocomm Research (I²R), A*STAR, Singapore

*Abstract*—This paper studies time difference of arrival (TDoA)-based algorithms for drone controller localization and analyzes TDoA estimation in multipath channels. Building on TDoA estimation, we propose two algorithms to enhance localization accuracy in multipath environments: the Maximum Likelihood (ML) algorithm, and the Least Squares Bancroft with Gauss-Newton (LS-BF-GN) algorithm. We evaluate these proposed algorithms in two typical outdoor channels: Wireless Local Area Network (WLAN) Channel F and the two-ray ground reflection (TRGR) channel. Our simulation results demonstrate that the ML and LS-BF-GN algorithms significantly outperform the LS-BF algorithm in multipath channels. To further enhance localization accuracy, we propose averaging multiple tentative location estimations. Additionally, we evaluate the impact of time synchronization errors among sensors on localization performance through simulation.

## I. INTRODUCTION

Unmanned aerial vehicles (UAVs) have become widely used in many applications such as aerial photography, agriculture, surveillance etc. However, the increase in the availability and accessibility of high-end UAVs has created new challenges in security. Unlawful usage of drones has caused threats to key infrastructures such as airports, power plants, and airbases. Most existing research on counter-drone approaches focuses on the detection and neutralization of the intruding drones [1]. However, it is equally important to detect and locate the drone controller to counteract the whole intrusion system. Detecting and locating the drone controller is complicated by the fact that there is no industrial standard for drone-controller communications. Different manufacturers choose their own proprietary communication protocols. Since mainstream high-end commercial drone controllers adopt frequency hopping spread spectrum (FHSS) technology to send commands to drones, we study the localization of FHSS based drone controllers.

For drone controller localization, the time difference of arrival (TDoA)-based method has been proposed in [2]. In this method, TDoA is measured at different receivers [3][4], requiring all receivers to be synchronous with high accuracy. The authors in [5] introduce an approach to localize the drone controller by using the angle of arrival (AoA) combined with triangulation. In [6], the authors propose using a deep learning approach to estimate AoA and triangulation to locate drone controllers. Both [5] and [6] require receivers equipped with antenna array to perform the AoA estimation [7]. In urban and suburban environments with rich multipath propagation and without dominate line of sight (LOS) path, traditional TDoA or AoA-based localization methods may fail to provide accurate results. To address localization accuracy issues in multipath channels, researchers have turned to RF fingerprint-based methods. RF Fingerprint-based methods have been widely used in indoor localization for the past decades. These methods utilize physical quantities such as the received signal strength indicator (RSSI) [8, 9], channel state information (CSI) [10, 11] or channel frequency response (CFR) [12]. The main advantage of RF fingerprint-based localization methods is that they do not require synchronization of all sensor nodes during data collection. However, these methods do require time-consuming data collection to build the training dataset.

In this paper, we study TDoA-based algorithm for drone controller localization and analyze the TDoA estimation in multipath channel. The TDoA-based techniques for localization are generally divided into two steps: firstly, the TDoA between sensors is estimated through cross correlation; secondly, the estimated TDoA is utilized to estimate the drone controller position. The conventional method for the second step is the Least-Square Bancroft (LS-BF) algorithm [18]. When the propagation channel from a drone controller to a sensor node is a multipath channel and without a dominant LOS path, traditional TDoA-based localization method cannot provide accurate results. To address this issue, we propose Maximum Likelihood (ML) algorithm and Least Square Bancroft with Gaussian Newton (LS-BF-GN). We evaluate our proposed algorithms in two typical outdoor channels: Wireless local area network (WLAN) channel F [19] and two ray ground reflection channels (TRGR) [20]. Our simulation results show that the proposed ML and LS-BF-GN algorithms achieve much better performance than traditional LS-BF algorithm in multipath channel. To further improve the localization accuracy, we propose to average multiple tentative location estimations. Since the TDoA-based localization method requires that all sensors are time-synchronous, we therefore evaluate the effect of time synchronization error among sensors on localization performance through simulation.

The remainder of this paper is organized as follows: In section II, we present our system model. In section III, we analyze the TDoA estimation in the multipath channel, which shows that multipath fading degrades the TDoA estimation. In section IV we present our proposed localization algorithms based on TDoA. In section V we present our simulation results. We conclude our work in section VI.

## II. SYSTEM MODEL

### A. Signal Model

Since most high-end drone controllers adopt the FHSS technologies, without loss of generality, we will use the FHSS signal model to describe the drone controller transmitted signal. As shown in Fig. 1, FHSS signal is pulsed, with each pulse owning its individual carrier frequency. Assume that pulse width and pulse period is $T_p$ and $T_0$, respectively. The number of pulses is assumed to be $N$. The $i$-th $(i = 1, 2, \cdots, N)$ pulse signal can be written as

$$s^i(t) = \begin{cases} a_i(t)e^{j(2\pi f_i t + \varphi_i)}, & (i-1)T_0 \leq t < (i-1)T_0 + T_p \\ 0, & (i-1)T_0 + T_p \leq t < iT_0 \end{cases} \quad (1)$$

where, $a_i(t)$, $f_i$, $\varphi_i$ denotes the baseband signal, carrier frequency and initial phase of the $i$-th pulse, respectively. The total signal $s(t)$ containing $N$ pulses can be written as,

$$s(t) = \sum_{i=1}^{N} s^i(t), \quad 0 \leq t < NT_0 \quad (2)$$

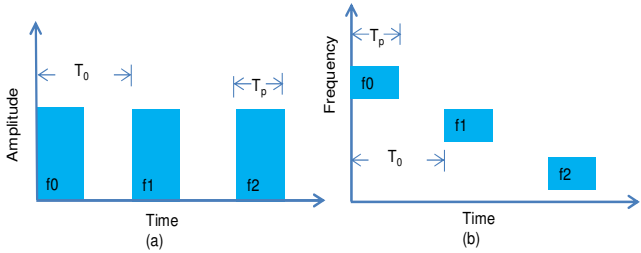

Fig. 1. FHSS signal (a) time domain waveform (b) Time-Frequency analysis

### B. Channel Model

For drone controller localization, we consider outdoor channel model. WLAN channel F [19] and TRGR channel [20] are two typical outdoor channels. WLAN channel F is a frequency selective fading channel with channel impulse response $h(\tau; t)$ being written as

$$h(\tau; t) = \sum_{l=0}^{L_h - 1} h_l \delta(\tau - \tau_l) \quad (3)$$

where, $L_h$ is the number of multipath, $h_l$ and $\tau_l$ are complex gain and delay of $l$-th path, respectively. $h_l$ is zero-mean complex-valued Gaussian random process. Different path gains are uncorrelated with respect to each other. According to [19], delay spread of WLAN channel F is 150 ns. When sampling rate is 160Msps, number of multipath in WLAN channel F is 169 taps.

The TRGR channel can be seen as a special case of the channel model in (3). The TRGR [20] channel is a propagation model in which the received signal has two components: the LOS component and the reflection component formed by a single ground-reflected wave [20].

## III. TDOA ESTIMATION IN MULTIPATH CHANNEL

The TDoA based localization techniques are generally divided into two steps. In the first step the TDoA between sensors are estimated through cross correlation. In the second step drone controller location is determined using estimated TDoA. In this section, we present the TDoA estimation in AWGN channel first, then we analyze the TDoA estimation in multipath channel, which shows that multipath fading degrades the TDoA estimation.

### A. TDoA estimation in AWGN channel

The cross correlation-based estimator is one of the most utilized estimators for TDoA estimation. In AWGN channel, the accuracy of TDoA estimation relies on the bandwidth of the signal. Without loss of generality, we assume number of sensors is $n$ and let sensor 1 as reference sensor for estimating TDoA. Assume that time delay between received signal of sensor 1 and received signal of sensor $j$ are $\tau_j$, the received signal of sensor 1 and sensor $j$ $(j = 2, \cdots, n)$, can be written as,

$$s_1(t) = \sum_{i=1}^{N} s^i(t) + w_1(t), \quad 0 \leq t < NT_0 \quad (4)$$

$$s_j(t) = \sum_{i=1}^{N} s^i(t - \tau_j) + w_j(t), \quad 0 \leq t < NT_0 \quad (5)$$

where, $w_1(t) \sim N(0, \sigma_1^2)$ and $w_j(t) \sim N(0, \sigma_j^2)$ are independent Gaussian noises.

The signal model for FH signal given by (2) is pulsed (not time-continuous), each pulse having different carrier frequency. Due to the above characteristics, the conventional CAF method can only use a single pulse. For the $i$-th pulse, the cross-ambiguity function (CAF) can be specified as [16],

$$CAF_{1,j}^i(\tau) = \int_{(i-1)T_0}^{(i-1)T_0 + T_p} s_1^*(t) s_j(t + \tau) dt \quad (6)$$

The maximum peak of $|CAF_{1,j}^i(\tau)|$ occurs at $\tau = \tau_j$. In AWGN channel, the estimation accuracy of TDoA using a single pulse is [16]

$$\sigma_t = \frac{1}{B_s \sqrt{B_n T_p \gamma}} \quad (7)$$

where, $B_s$ denotes the root mean square bandwidth of signal, $B_n$ denotes the noise bandwidth and

$$\frac{1}{\gamma} = \frac{1}{2} \left( \frac{1}{\gamma_1} + \frac{1}{\gamma_j} + \frac{1}{\gamma_1 \gamma_j} \right) \quad (8)$$

with $\gamma_1$ and $\gamma_j$ denoting SNR of $s_1(t)$ and $s_j(t)$, respectively. Since $T_p$ is small, i.e., in the order of ms, the estimation accuracy for TDoA is limited.

Since hopping frequency of each pulse is different, the CAFs cannot be coherently accumulated using the method in [10]. The estimation accuracy of TDoA can be improved by using incoherent accumulation, which can be written as [16],

$$\hat{\tau}_{1,j} = \frac{1}{N} \sum_{i=1}^{N} \hat{\tau}_{1,j}^i \quad (9)$$

where, $\hat{\tau}_{1,j}^i$ denotes the estimated TDoA based on $CAF_{1,j}^i(\tau)$ in (6).

## B. TDoA estimation in multipath channel

In multipath channel, the received signal of sensor 1 and sensor $j$ ($j = 2, \cdots, n$) can be written as,

$$y_1(t) = \sum_{l=0}^{L_h-1} s(t - \tau_{l,1}) h_{l,1} + w_1(t) \quad (10)$$

$$y_j(t) = \sum_{k=0}^{L_h-1} s(t - \epsilon_{k,j} - \tau_j) g_{k,j} + w_j(t) \quad (11)$$

where, $h_{l,1}$ and $\tau_{l,1}$ are time-varying complex gain and delay of $l$-th path of propagation channel from drone controller to sensor 1, and $g_{k,j}$ and $\epsilon_{k,j}$ are time-varying complex gain and delay of $k$-th path of propagation channel from drone controller to sensor $j$. Correspondingly, the CAF function can be written as,

$$\overline{CAF}_{1,j}^i(\tau) = \int_{(i-1)T_0}^{(i-1)T_0 + T_p} y_1^*(t) y_j(t + \tau) dt$$

$$= \int_{(i-1)T_0}^{(i-1)T_0 + T_p} \sum_{l=0}^{L_h-1} \sum_{k=0}^{L_h-1} s^*(t - \tau_{l,1}) h_{l,1}^* s(t - \epsilon_{k,j} - \tau_j + \tau) g_{k,j} dt$$

$$= \sum_{l=0}^{L_h-1} \sum_{k=0}^{L_h-1} \int_{(i-1)T_0}^{(i-1)T_0 + T_p} s^*(t - \tau_{l,1}) s(t - \epsilon_{k,j} - \tau_j + \tau) h_{l,1}^* g_{k,j} dt$$

$$= \sum_{l=0}^{L_h-1} \sum_{k=0}^{L_h-1} A_{k,l}(\tau) \quad (12)$$

where, $A_{k,l}(\tau) = \int_{(i-1)T_0}^{(i-1)T_0 + T_p} s^*(t - \tau_{l,1}) s(t - \epsilon_{k,j} - \tau_j + \tau) h_{l,1}^* g_{k,j} dt$.

There are $(L_h)^2$ different items of $A_{k,l}(\tau)$ ($0 \leq l \leq L_h - 1, 0 \leq k \leq L_h - 1$). Without loss of generality, we assume that the relative delay of first path is zero, i.e., $\tau_{0,1} = 0$ and $\epsilon_{0,j} = 0$, then $A_{0,0}(\tau)$ will achieve maximal value when $\tau = \tau_j$. For $k > 0$ and $l > 0$ $|A_{k,l}(\tau)|$ will achieve a maximal value when $\tau$ is different from $\tau_j$. If first path is not dominant, from (12) we can see that due to $L_h \gg 1$, the maximal peak value of $\left|\overline{CAF}_{1,j}^i(\tau)\right|$ no longer occurs at $\tau = \tau_j$, thus the performance of TDoA estimation is degraded in multipath channel.

Like AWGN channel scenario, the estimation accuracy of TDoA in multipath channel can be improved by using incoherent accumulation,

$$\hat{\tau}_{1,j} = \frac{1}{N} \sum_{i=1}^{N} \bar{\tau}_{1,j}^i \quad (13)$$

In this case, $\bar{\tau}_{1,j}^i$ denotes the estimated TDoA based on $\overline{CAF}_{1,j}^i(\tau)$ in (12).

## IV. PROPOSED METHOD FOR DRONE CONTROLLER LOCALIZATION

In this section, we present our proposed method for drone controller localization. With TDoA estimation being obtained, the conventional method for estimating the drone controller location is the LS-BF algorithm [18]. When propagation channel from drone controller to sensor node is multipath channel without a dominate LOS path, conventional LS-BF algorithm cannot provide accurate localization performance. To address this issue, we propose two algorithms to improve the localization accuracy of drone controller.

- Maximum likelihood (ML)
- Least Square Bancroft (LS-BF) with Gaussian Newton

### A. Conventional Least Square Bancrof (LS-BF) Algorithm

Assume $v_j$ is the coordinates of the $j$-th sensor ($j = 1, \cdots, n$) and $x$ is the coordinates of the drone controller, and $r_j$ is the distance between drone controller and sensor $j$. The squared range equation can be expressed as

$$r_j^2 = ||v_j||^2 - 2v_j^T x + ||x||^2 \quad (14)$$

According to [18], the two possible estimation of drone controller location is

$$\hat{x}_l = p t_l + q, \quad l = 1,2 \quad (15)$$

where $p = (A^T A)^{-1} A^T \mathbf{1}$ and $q = (A^T A)^{-1} A^T b$, and $A$ and $b$ can be expressed as,

$$A = \begin{bmatrix} 2v_1^T \\ \ldots \\ 2v_n^T \end{bmatrix}, \quad b = \begin{bmatrix} ||v_1||^2 - r_1^2 \\ \ldots \\ ||v_n||^2 - r_n^2 \end{bmatrix} \quad (16)$$

$t_1$ and $t_2$ are two solutions to the following equation:

$$||\mathbf{p}||^2 t^2 + (2p^T q - 1) t + ||q||^2 = 0 \quad (17)$$

The solution which leads to the smaller error with TDoA estimation is selected as the final solution.

### B. Proposed ML based localization method

Our proposed ML algorithm can be described as follows:

**Step 1**: Calculate cross ambiguity function (CAF), find maximal value of CAF and obtain TDoA between sensor $j$ and sensor 1 $\hat{\tau}_{1,j}$ ($j = 2, \ldots, n$), according to (13).

**Step 2**: Divide the whole interested area into $M$ small grid, denoted $z_k$ as the coordinates of the $k$-th grid, $k = 1, 2, \ldots, M$.

**Step 3**: For $k$-th grid, calculate the TDoA between sensor 1 and sensor $j$ as follows,

$$\tilde{\tau}_{1,j}(k) = (\| z_k - v_1 \| - \| z_k - v_j \|) / c, \, j = 2, \ldots, n \quad (18)$$

where, $v_i$ ($i = 2, \ldots, n$) is coordinates of sensor $i$, and $c$ is the light of speed. The error between $\tilde{\tau}_{1,j}(k)$ and TDoA estimation $\hat{\tau}_{1,j}$ can be expressed as,

$$e_k = \sum_{j=2}^{n} (\tilde{\tau}_{1,j}(k) - \hat{\tau}_{1,j})^2, \quad k = 1, 2, \ldots, M \quad (19)$$

**Step 4**: Choose the grid which has the smallest error with the TDoA estimation $\hat{\tau}_{1,j}$,

$$J = \arg\min_k e_k = \arg\min_k \sum_{j=2}^{n} (\tilde{\tau}_{1,j}(k) - \hat{\tau}_{1,j})^2 \quad (20)$$

The ML estimation of drone controller location is $\hat{x}_{ML} = z_J$.

## C. Proposed Least Square Bancrof with Gaussian Newton (LS-BF-GN)

Gaussian Newton method is an iterative approach for solving nonlinear equations, which needs an initial estimate to ensure globe convergence. In our proposed LS-BF-GN algorithm, we use LS-BF algorithm to obtain the tentative drone controller location $\hat{x}_{LS-BF}$, which is used as the initial estimate for Gaussian Newton method.

Assume $v_j$ is the coordinates of the $j$-th sensor ($j = 1, \cdots, n$) and $x$ is the coordinates of the drone controller, our proposed LS-BF-GN can be described as follows:

**Step 1**: Calculate cross ambiguity function (CAF), find maximal value of CAF and obtain TDoA between sensor $j$ and sensor 1 $\hat{\tau}_{1,j}$ ($j = 2, \ldots, n$), according to (13).

**Step 2:** Based on TDoA estimation $\hat{\tau}_{1,j}$ ($j = 2, \ldots, n$), use LS-BF algorithm to estimate tentative drone controller location $\hat{x}_{LS-BF}$ and assign $\lambda_1 = \hat{x}_{LS-BF}$.

**Step 3**: Choose initial guess $x_0$ equals to $\lambda_1$ and stopping tolerance $\delta$, set k = 0

**Step 4** : Compute

$$J_k(x) = \begin{bmatrix} \frac{(v_2 - v_1)^T}{\| v_2 - v_1 \|} \\ \vdots \\ \frac{(v_n - v_1)^T}{\| v_n - v_1 \|} \end{bmatrix} \quad (21)$$

set $x_{k+1} = x_k + \Delta x_k$, where, $\Delta x_k = J_k(x) \backslash (h(x_k) - u)$, and $x = A \backslash b$ means let $x$ be the least square solution to $Ax = b$. $h(x_k)$ and $u$ can be written as,

$$h(x_k) = \begin{bmatrix} \| v_2 - x_k \| - \| v_1 - x_k \| \\ \vdots \\ \| v_n - x_k \| - \| v_1 - x_k \| \end{bmatrix} \quad (22)$$

$$u = \begin{bmatrix} \hat{\tau}(2,1)c \\ \vdots \\ \hat{\tau}(n,1)c \end{bmatrix} \quad (23)$$

**Step 5**: If stopping condition $\| \Delta x_k \| < \delta$ is not satisfied and $k \leq k_{max}$, increment $k$ and repeat from step 4.

**Step 6:** Denotes the iteratively refined location estimation as $\lambda_2$.

**Step 7:** Final location estimation is chosen between $\lambda_1$ and $\lambda_2$, which ever has less error compared with TDoA estimation $\hat{\tau}(i,1)$ ($i = 2, \ldots, n$).

$$\hat{x} = \begin{cases} \lambda_1 \text{ if } a < \tilde{a} \\ \lambda_2, \text{ if } a > \tilde{a} \end{cases} \quad (24)$$

where, $a = \sum_{i=2}^{n}(\zeta_{1,i} - \hat{\tau}(i,1))^2$, $\tilde{a} = \sum_{i=2}^{n}(\tilde{\zeta}_{1,i} - \hat{\tau}(i,1))^2$ and
$\zeta_{1,i} = (\| \lambda_1 - v_1 \| - \| \lambda_1 - v_i \|)/c$, $i = 2, \ldots, n$,
$\tilde{\zeta}_{1,i} = (\| \lambda_2 - v_1 \| - \| \lambda_2 - v_i \|)/c$, $i = 2, \ldots, n$

## D. Averaging a number of tentative location estimation to improve the localization performance

During our study, we found that in multipath rich channel and without a dominate LOS path, even using our proposed ML algorithm and LS-BF-GN algorithm, the localization performance is still not satisfied. Thus, we propose to average multiple tentative location estimations to improve the localization performance. Note that since drone controllers are relatively static, thus it is valid to average the detected location over a relatively long period. Assume that for a given drone controller location, TDoA estimation is performed $N_{avg}$ times, using $N$ pulses for each TDoA estimation. Based on each TDoA estimation, we estimate the drone controller location and denote $j$-th location estimation as $\hat{x}(j)$. Then final drone controller location is the average of $N_{avg}$ location estimation,

$$\tilde{x} = \frac{1}{N_{avg}} \sum_{j=1}^{N_{avg}} \hat{x}(j) \quad (25)$$

Note that, during the multiple TDoA and location estimation, the channel gain and delay will change with time.

## V. SIMULATION EVALUATION

In this section, we evaluate the localization performance through simulation. We adopt 2FSK (Frequency shift keying) as baseband modulation to generate drone controller transmitted signal. The generated drone controller signal has a pulse period of $T_0 = 8 \ ms$ and a pulse width of $T_p = 2 \ ms$. In our simulation of WLAN channel F, the values of $h_l$ and $\tau_l$ remain constant for a duration of 80 $ms$, then changed to a new set of values that are independent of previous ones. We use $N = 10$ pulses for each TDoA estimation. The hopping frequency of drone controller is uniformly distributed in [2400 MHz, 2480 MHz]. Our simulation scenario is a 50 $m$ by 50 $m$ square area, with four sensors put on the four corners. The drone controller position is uniformly distributed in the square area. We evaluate the drone controller localization performance in both TRGR channel and WLAN channel F.

### A. Localization performance in TRGR channel

Since TDoA based localization method requires all sensors to be time-synchronized, it is crucial to assess the impact of time synchronization errors on localization performance. When accounting for synchronization errors among sensors, the actual TDoA estimation between sensor 1 and sensor $j$ can be written as,

$$\hat{\tau}_{1,j} = \frac{1}{N} \sum_{i=1}^{N} \bar{\tau}_{1,j}^i + \delta_j \quad (26)$$

where, $\bar{\tau}_{1,j}^i$ denotes the estimated TDoA based on $\overline{CAF}_{1,j}^i(\tau)$, $\delta_j$ is the time synchronization error between sensor 1 and sensor $j$, which is a zero mean random variable uniformly distributed in $[-\alpha, \alpha]$.

Fig.2 shows the effect of time synchronization error on the localization performance in TRGR channel. In this figure, number of pulses used for TDoA estimation is 10 ($N = 10$). averaging is not used (i.e., $N_{avg} = 1$). Time synchronization error $\delta_j$ between sensor 1 and sensor $j$ ($j$ = 2, 3, 4) is uniformly distributed in the range $[-\alpha, \alpha]$, with $\alpha$ taking the values of 0 $ns$, 10 $ns$, 20 $ns$. We can see that our proposed ML algo and LS-BF-GN achieve much better performance than that of LS-BF algo. When $\alpha$ is 0 $ns$, the 90[th] percentile localization error is 1 $m$, 1 $m$ and 1.8 $m$ for ML algo, LS-BF-GN algo and LS-BF algo, respectively. When $\alpha$ is 10 $ns$, the 90[th] percentile localization error is 2.5 $m$, 2.8 $m$ and 6 $m$ for ML algo, LS-BF-GN algo and LS-BF algo, respectively. When $\alpha$ is 20 $ns$, the 90[th] percentile localization error is 4.4 $m$, 5.8 $m$ and 11 $m$ for ML algo, LS-BF-GN algo and LS-BF algo, respectively.

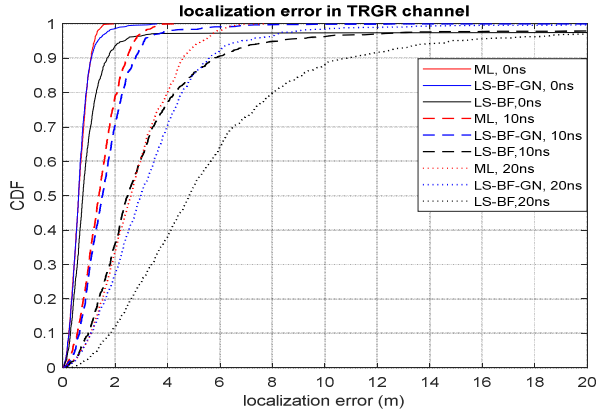

Fig. 2. Effect of time synchronizaion error on localization performance in TRGR channel.

Fig.3 shows the effect of average number $N_{avg}$ on localization performance in TRGR channel. In this figure, $\alpha$ is 20 $ns$. From this figure we can see that without averaging ($N_{avg} = 1$), the 90[th] percentile localization error is 4.6 $m$ 5.5 $m$ and 11.7 $m$ for ML algo, LS-BF-GN algo and LS-BF algo, respectively. However, when $N_{avg} = 20$, the 90[th] percentile localization error is improved to 1.49 $m$, 1.87 $m$ and 6.55 $m$ for ML algo, LS-BF-GN algo and LS-BF algo, respectively. Thus we can see using averaging can greatly improve the localization performance.

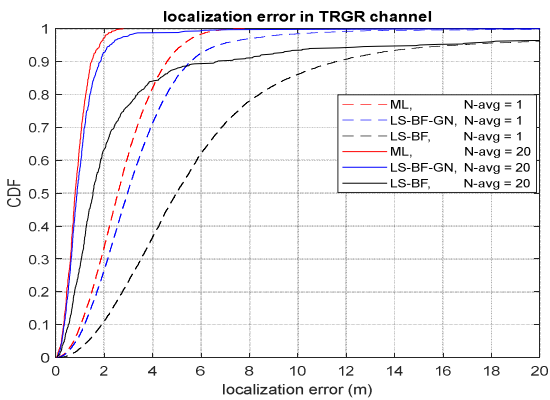

Fig. 3. The effect of average number on localization performance in TRGR channel

Fig.4 shows the TDoA estimation error in TRGR channel. In this figure, $\alpha$ is 0 $ns$. Since TRGR chanle has dominate LOS path, the TDoA estimation error is within [-8 $ns$, 8 $ns$], which explain the good localization shown in Fig.2 when $\alpha = 0$ $ns$.

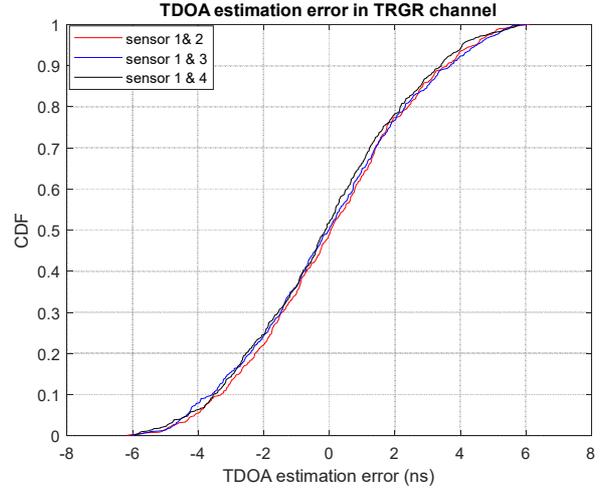

Fig. 4. TDoA estimation performance in TRGR channel

### B. *Localization performance in WLAN channel F*

Fig.5 shows the effect of averaging on localization performance in WLAN channel F. Since WLAN channel F has large number of multipath without dominant LOS path, the localizatin performance is much worse than that of TRGR channel. We can see that our proposed ML algo and LS-BF-GN achieve much better performance than that of LS-BF algo. Without averaging ($N_{avg} = 1$), the 90[th] percentile localization error is 21.5 $m$, 26 $m$ and 55 $m$ for ML algo, LS-BF-GN algo and LS-BF algo, respectively. With averaging ($N_{avg} = 20$) the localization performance is improved greatly in multipath channel (without LOS path). With $N_{avg} = 20$, the 90[th] percentile localization error is 10 $m$, 10 $m$ and 15 $m$ for ML algo, LS-BF-GN algo and LS-BF algo, respectively.

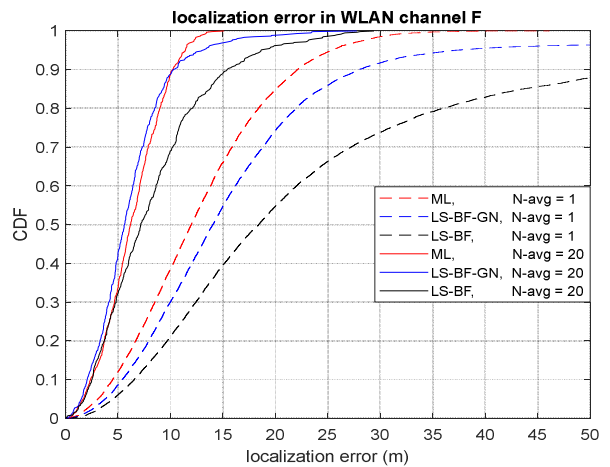

Fig. 5. Effect of averaging on localizationpPerformance in WLAN channel F

Fig.6 shows the TDoA estimation error in WLAN channel F. In this figure, $\alpha$ is 0 $ns$. Since WLAN channel F has large number of multipath without LOS path, the TDoA estimation performance is much worse than that of TRGR channel. We can see that TDoA estimation error is within [-300 $ns$, 300 $ns$], which can explain the worse localization shown in Fig.5 when $N_{avg} = 1$.

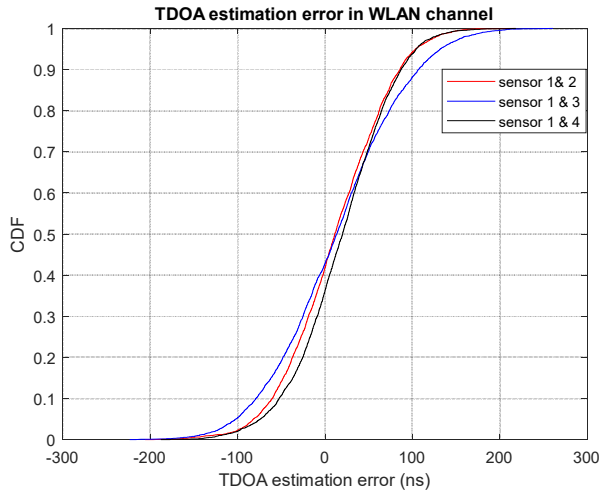

Fig. 6.  TDoA estimation performance in WLAN channel F

## VI. CONCLUSION

We have studied TDoA based algorithm for drone controller localization. We propose two algorithms to improve localization accuracy in multipath channel: Maximum Likelihood (ML) algorithm and Least Square Bancroft with Gaussian Newton (LS-BF-GN). We evaluate our proposed algorithms in WLAN channel F and TRGR channel. Our simulation results show that the performance of proposed ML and LS-BF-GN algorithm achieve much better performance than traditional LS-BF algorithm. To further improve the localization accuracy, we propose to average multiple tentative location estimations. We evaluate the effect of time synchronization error among sensors on the localization performance in TRGR channel.

### ACKNOWLEDGEMENT

This research is supported by the National Research Foundation, Singapore and Infocomm Media Development Authority under its Future Communications Research Development Programme.

### REFERENCES

[1] M. M. Azari, H. Sallouha, A Chiumento, S. Rajendran, E. Vinogradov, and S. Pollin, Key Technologies and System Trade-offs for Detection and Localization of Amateur Drones, IEEE Communications Magazine, January 2018.

[2] B. P. Teh, RF techniques for detection, classification and location of commercial drone controllers, https://tekmarkgroup.com/eshop/image/catalog/Application/AEROSPACE/Paper-5_Techniques-forDetection-Location-of-Commercial-Drone-Controllers_2017-MalaysiaAD-Symposium.pdf, Tech. Rep. July 2017.

[3] G. Shen, R. Zetik, and R. S. Thomä, Performance Comparison of TOA and TDOA Based Location Estimation Algorithms in LOS Environment, Proceeding of the 5th workshop on Poistioning , Navigation and communication 2008 (WPNC'08).

[4] H. Chen, T. Ballal, et al. A joint TDOA-FDOA localization approach using particle swarm optimization." IEEE Wireless Communications Letters, Vol. 9, No.8, 2020.

[5] P. Nguyen, T. Kim,  J. Miao and D. Hesselius,  Towards RF-based Localization of a Drone and Its Controller, DroNet 19, ACM, June 2019.

[6] D. Shorten, S. Srivastava and J. Murray, Localisation of Drone Controllers from RF Signals using a Deep Learning Approach, PRAI Aug 2018 (Proceedings of the International Conference on Pattern Recognition and Artificial Intelligence).

[7] S. Basak, and B. Scheers. Passive radio system for real-time drone detection and DoA estimation, 2018 International Conference on Military Communications and Information Systems (ICMCIS). IEEE, 2018.

[8] P. Bahl and V. Padmanabhan. RADAR: An inbuilding RF-based user location and tracking system. In Proc. of IEEE Infocom, pages 775–784, 2000.

[9] M. Youssef and A. Agrawala, The Horus WLAN Location Determination, MobiSys 2005.

[10] K. Wu, J. Xiao, Y. Yi, D. Chen, X. Luo and L. M. Ni, CSI-Based Indoor Localization, IEEE Trans. on Parallel and Distributed Systems, Vol.24, No.7, July 2013.

[11] X. Wang, L. Gao, S. Mao and S. Pandey, DeepFi: Deep Learning for Indoor Fingerprinting Using Channel State Information, WCNC 2015.

[12] X. Wang, X. Wang, S. Mao, Deep Convolutional Neural Networks for Indoor Localization with CSI Images, IEEE Transactions on Network Science and Engineering, Vol. 7, No. 1, 2020.

[13] C. H. Knapp and G. C. Caprter, The generalized correlation method for estimation of time delay, IEEE Transactions on Acoustics, Speech, and Signal Processing, Vol.24, No.4, 1976.

[14]  J. Wang, Y. Xu, and P. Xu, Linear method for TDOA estimation of frequency-hopping signal, 2012 8th International Conference on Wireless Communications, Networking and Mobile Computing, IEEE, 2012.

[15] X. Ouyang, Q. Wan, J. Cao, J. Xiong and Q. He, Direct TDOA geolocation of multiple frequency-hopping emitters in flat fading channels, IET Signal Processing, Vol.11, No.1, 2016.

[16] D. Hu, Z. Huang, K. Liang, S. Zhang, Coherent TDOA/FDOA estimation method for frequency-hopping signal, 2016 8th International Conference on Wireless Communications & Signal Processing (WCSP). IEEE, 2016.

[17] X. Ouyang, Q. He, Y. Yang and Q. Wan,  TDOA/FDOA estimation algorithm of frequency-hopping signals based on CAF coherent integration, IET Communications Vol.14, No.2, 2019, pp. 331-336.

[18] N. Sirola, Closed-form algorithms in mobile positioning: Myths and misconceptions, IEEE 7th  Workshop on Positioning, Navigation and Communication (WPNC), 2010.

[19] V. Erceg, et.al., TGn Channel Models, IEEE 802.11-03/940r4, May, 2004.

[20] C. C. Chiu, A. H. Tsai, H. P. Lin, C. Y. Lee, and L. C. Wang, Channel Modeling of Air-to-Ground Signal Measurement with Two-Ray Ground-Reflection Model for UAV Communication Systems, 2021 30th Wireless and Optical Communications Conference (WOCC), 2021.

[21] S. N. A. Ahmed and Y. Zeng, UWB Positioning Accuracy and Enhancements, TENCON 2017 Malaysia.